 \documentclass[pra,twocolumn,lengthcheck,superscriptaddress,showpacs,preprintnumbers,amsmath,amssymb]{revtex4}

 \usepackage{graphicx}
 \usepackage{dcolumn}
 \usepackage{bm}

\begin{document}

 \title{Fractional-filling Mott domains in two dimensional optical superlattices }

 \author{P. Buonsante}
 \affiliation{Dipartimento di Fisica, Politecnico di Torino and I.N.F.M, Corso Duca degli Abruzzi 24, I-10129 Torino (ITALIA)}%
 \author{V. Penna}
 \affiliation{Dipartimento di Fisica, Politecnico di Torino and I.N.F.M, Corso Duca degli Abruzzi 24, I-10129 Torino (ITALIA)}%
 \author{A. Vezzani}
 \affiliation{Dipartimento di Fisica, Universit\`a degli Studi di Parma, I-43100 Parma (ITALIA)}

 \date{\today}

 \begin{abstract}
 Ultracold bosons in optical superlattices are expected to exhibit fractional-filling insulating phases for sufficiently large repulsive interactions. On strictly 1D systems, the exact mapping between hard-core bosons and free spinless fermions shows that any periodic modulation in the lattice parameters causes the presence of  fractional-filling insulator domains. Here, we focus on two recently proposed realistic 2D structures where such mapping does not hold, i.e. the two-leg ladder and the trimerized kagom\'e lattice. Based on a {\it cell} strong-coupling perturbation technique,  we provide quantitatively satisfactory phase diagrams for these structures, and give estimates for the occurrence of the fractional-filling insulator domains in terms of the inter-cell/intra-cell hopping amplitude ratio. 
 \end{abstract}

 \pacs{
 05.30.Jp,  
 73.43.Nq,  
 03.75.Lm 
 74.81.Fa,   
 }

 \maketitle


 Originally introduced as a toy model of liquid helium trapped in porous media \cite{A:Fisher}, the Bose Hubbard model is nowadays routinely brought to experimental reality in terms of ultracold bosonic atoms trapped in optical or magnetic lattices \cite{A:Jaksch}. The power of this physics tool as a virtually ideal realization of the theoretical model stands in the broad range of parameters and configurations attainable. For instance, the hopping amplitude of the bosons across the lattice sites can be controlled quite directly by varying the strength of the laser beams providing the optical confinement. 
 Additional flexibility comes from the possibility of tuning --- and even reversing --- the interparticle interactions via Feshbach resonances \cite{A:Theis}. Further aspects of ideality are the fact that, to all practical purposes, the system is at zero temperature and isolated from the environment. Thus, for instance,  the phononic excitations  typical of realistic condensed matter lattice systems are ruled out. This impressive degree of control over the system parameters played a key role in the breakthrough experiment in which the superfluid-insulator quantum phase transition predicted for the Bose-Hubbard model \cite{A:Fisher} was actually observed for a gas of ultracold atoms trapped in an optical lattice \cite{A:Greiner02}.

 A great variety of optical lattices  can be attained through a suitable choice of the number and setup of the laser beams providing the optic confinement \cite{A:Blakie}. So far, 1D \cite{A:TrombS}, 2D \cite{A:Greiner}, and 3D \cite{A:Greiner02} Euclidan lattices have been realized, as well as 1D superlattices \cite{A:Peil} and quasiperiodic lattices \cite{A:Guidoni97}.
 The progress in optical trapping techniques prompted a great number of proposals for non trivial optical lattices, including periodic geometries \cite{A:Amico05}, quasicrystals \cite{A:Sanchez,A:Salerno} and 2D superlattices \cite{A:Santos}.

 The complex periodic geometry of optical superlattices, entailing a multi-band single-particle spectrum, allows for fractional filling insulating domains in the zero temperature phase diagram of superlattice BH models \cite{A:Motrunich,A:Santos,A:Roth03,A:LobiMF}. It has been furthermore observed that such domains may exhibit an unusual loophole shape, as opposed to the roughly triangular shape of the customary Mott lobes \cite{A:loophole}.
  So far, the theoretical investigations on the quantum phase transitions in superlattice BH models focused on 1D systems, where the fractional filling insulating domains can be directly related to the inter-band gaps in the single particle spectrum \cite{A:loophole}, owing to the mapping between the hard-core boson regime and free spinless fermions \cite{A:Girardeau,A:Cazalilla}.
 On non strictly 1D structures such mapping does not apply, and therefore the phase diagram cannot be inferred from the single particle spectrum.
 Qualitative results have been obtained in the extreme regime where the hopping amplitudes among different cells are much smaller than those within the same cell \cite{A:Santos}.
 Quantitative results can be obtained  resorting to  numerical methods such as quantum Monte Carlo \cite{A:Batrouni,A:Kashurnikov96} or density matrix renormalization group algorithms \cite{A:Kuehner}. These approaches necessarily address finite systems, whose size has to be sufficiently large to provide a good approximation of the thermodynamic limit. On superlattices the dimensional scaling of the computational demand is made even more serious by the fact that the building blocks  are cells comprising several sites, as opposed to single sites.  On the other hand, satisfactory quantitative results have been obtained for $d$-dimensional regular lattices based on analytical \cite{A:Freericks1} and  numeric \cite{A:Elstner99a} strong-coupling perturbative expansions (SCPE). A similar technique, adapted to the complex periodicity of superlattices \cite{A:cSCPE}, has been recently introduced for the study of the 1D structures whose experimental realization is described in Ref.~\cite{A:Peil}.
 \begin{figure}
 \begin{center}
 \includegraphics[width=8cm]{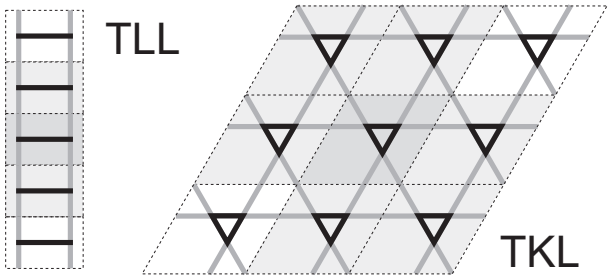}
 \caption{\label{F:lattices} Cell structure of the TLL (left) and TKL lattice (right). Dark and light thick solid lines represent intra-cell and inter-cell hopping amplitudes, respectively. A light gray shade signals the nearest neighbors of the central cell, which has a darker background. This shows that cell-lattice of the TLL (TKL) is a  1D euclidean lattice (2D triangular lattice).  }
 \end{center}
 \end{figure}

 Here we extend this technique, called {\it cell} SCPE, to more complex  superlattices. We focus on the two realistic structures sketched in Fig.~\ref{F:lattices}, namely the trimerized kagom\'e lattice (TKL) proposed in Ref.~\cite{A:Santos} and a two-leg ladder (TLL), which in some sense represents the minimal diversion from strictly 1D systems. Such a structure can be realistically obtained by superimposing a directional confinement on a suitably chosen portion of a 2D optical lattice, as it was done in Ref.~\cite{A:Peil} in the case of a 1D lattice, or combining simple lattices with different lattice constants, as proposed in Ref.~\cite{A:Garcia}.
 We observe that the strong interaction regime  of the ladder BH Hamiltonian maps onto a spin ladder model, a system that has received wide attention in the literature \cite{A:Dagotto,A:Vekua}.
This once again shows that ultracold atoms in optical lattices provide a direct realization of condensed matter models over a wide range of model parameters \cite{A:Garcia}.

As expected, we find that the structures we consider exhibit insulating domains at critical fractional fillings of the form $k/\ell$, where $k$ is a positive integer and $\ell$ is the number of sites in the unit cell ($\ell=2$ for the TLL and $\ell=3$ for the TKL). Since we assume that there are no energy offsets between the lattice sites, the non integer fillings correspond to loophole domains \cite{A:loophole}. Our aim is to investigate the conditions for the occurrence of such domains. Before getting into details, let us briefly list our main results.
Unlike the 1D case, on the TKL the superlattice structure is not sufficient for the occurrence of fractional-filling domains. That is to say, the inter-cell hopping amplitude has to be sufficiently smaller than the intra-cell amplitude for these insulating domains to occur, despite the fact that an arbitrarily small difference between these amplitudes is sufficient for opening a gap in the single-particle spectrum of the TKL.
Conversely, on the TLL our results suggest that  fractional filling insulating phases occur even if the single particle spectrum has no gap.
These discrepancies are consistent with the fact that on non strictly 1D structures the hard core limit of the Bose-Hubbard model is not equivalent to a free spinless fermion model \cite{A:Girardeau}.

Let us now introduce the notation for the superlattice BH Hamiltonian describing the optically trapped ultracold boson systems we are interested into.
The  spatial arrangement of the unit cells inherent in a superlattice is described by the so-called adjacency matrix $A$ of the cell-lattice, where $A_{c c'}$ is 1 if the cells labeled $c$ and $c'$ are adjacent, and zero otherwise.
As it is clear from Fig.~\ref{F:lattices}, the cell lattice is a 1D chain for the TLL, and a  triangular lattice for the TKL. 
The homogeneity of the cell lattice entails that the coordination of the generic cell $c$ is actually cell independent, $z_c =  \sum_c' A_{c c'} = z$.
A generic site of the superlattice can be labeled with two indices, 
the first referring to the cell it belongs to, and the second denoting
its position within such cell. Thus, the hopping amplitudes across
sites belonging to adjacent $\ell$-site cells can be described introducing a
 set of $\ell \times \ell$ matrices $t^{c c'}$. The row index of such
matrices refers to the cell denoted by the first superscript, whereas
the column index refers to the cell relevant to the second
superscript. Symmetry considerations lead to conclude that there are
$z/2$ such matrices, and that $t^{c' c} = (t^{c c'})^{\rm t}$. For
instance, in the case of the TLL, where $z = 2$, one has only
$t^{c, c+1}_{i j}  =t \delta_{i j}$. 
A further $\ell \times \ell$, symmetric, matrix $T_{j h}$ allows the description of the hopping amplitudes among sites belonging to the same unit cell. Thus, the BH Hamiltonian on a generic superlattice reads
\begin{equation}
\label{E:BH}
H = H_0 + \tau V, \quad H_0 = \sum_c H_c 
\end{equation}
where, introducing the the boson operators at the $j^{\rm th}$ site of the $c^{\rm th}$ cell, $a_{c,j}$, $a^+_{c,j}$ and $n_{c,j} = a^+_{c,j} a_{c,j}$,
\begin{eqnarray}
\label{E:cBH}
H_c &=& \sum_{j=1}^\ell \left[ \frac{U}{2} n_{c,j} (n_{c,j} -1)- (\mu-v_j) n_{c,j}\right] \nonumber \\ 
&-& \sum_{j,h =1}^\ell T_{h j} a^+_{c,j} a_{c,h}
\end{eqnarray}
refers to an isolated cell, while
\begin{equation}
\label{E:hop}
V = \sum_{c,c'} A_{c c'} \sum_{j,j'=1}^\ell t^{c c'}_{j j'} a^+_{c,j} a_{c',j'}
\end{equation}
takes into account the interaction among adjacent cells.
The parameters $U>0$, $\mu$ and $v_j$ appearing in Eq.~(\ref{E:cBH}) are the boson-boson (repulsive) interaction, the chemical potential and the energy offset of  the $j^{\rm th}$ site within the unit cell \cite{N:ovvio}.
We observe that the partitioning of a superlattice into a homogeneous lattice of identical unit cells, and hence the arrangement of terms in Eq.~(\ref{E:BH}), is not unique. However, the perturbative approach we are interested into suggests that a convenient choice is such that the inter-cell amplitudes are smaller than the intra-cell amplitudes.

The zero temperature phase diagram of the superlattice BH model in
Eq.(\ref{E:BH}) is standardly obtained in terms of the ground state
energies relevant to different fillings \cite{A:Fisher}. We recall indeed that
Hamiltonian $H$ commutes with the total number operator, $\sum_{c,j} n_{c,j}$, and hence it can be studied within a fixed-number sector of the
Fock space without loss of generality. Denoting by $E_N$ the ground state 
energy of $H$ relevant to a population of $N$ bosons, the boundaries
of the insulator domain possibly corresponding to a given filling $f$ are
standardly obtained as $\mu_\pm = \pm (E_{f M \pm 1} -E_{f M})$, where
$M$ is the number of sites in the lattice. The insulator domain
actually exists if the inequality $\mu_+> \mu_-$ holds strictly \cite{A:Fisher}.

As we mention above, we provide the insulator domain boundaries
$\mu_\pm$ in terms of strong coupling expansions for the ground state
energy of Hamiltonian (\ref{E:BH}), assuming that the inter-cell
hopping term in Eq. (\ref{E:hop}) is the perturbative quantity.
As it is easily understood, the eigenvectors and eigenvalues of the 
(identical) cell Hamiltonians (\ref{E:cBH}) are the key quantities
in our perturbative expansions. Denoting by $|k,N\rangle$ the  $k_c^{\rm th}$
eigenstate of the cell Hamiltonian relevant to a population of $N$
bosons and $E_k^N$ the corresponding eigenvalue, the generic
eigenstate of the unperturbed Hamiltonian $H_0$ relevant to a 
population of $\sum_c N_c$ bosons and the corresponding energy
can be written as
\begin{equation}
\label{E:unp}
|{\bf k}, {\bf N}\rangle\rangle = \bigotimes_c |k_c,N_c\rangle_c,
 \quad E({\bf k}, {\bf N}) = \sum_c E_{k_c}^{N_c},
\end{equation}
where the subscript $c$ labels the cells. In particular, setting $k_c = 1$ and $N_c = L$, Eq. (\ref{E:unp}) describes the unperturbed
ground state relevant to the fractional filling $f = L/\ell$, where we
recall that $\ell$ denotes the number of sites in each of the identical
cells. As we mention above, in order to describe the insulator domain
boundaries for such filling, we also need the ground state energies
for the so called {\it defect} states \cite{A:Freericks1}, obtained
increasing or decreasing the total population by a single boson. 
Equation (\ref{E:unp}) gives the correct unperturbed result for such
states, provided that the population of one of the cells is increased
or decreased by one boson. However, such energy is clearly degenerate,
since the population variation can affect any cell. This means that
the defect states must be treated according to degenerate perturbation
theory. We carried out the expansions up to the second order for both
the TLL and the TKL, obtaining the ground state
energies  for the fractional fillings $f = L/\ell$ and the relevant 
defect states in terms of the unperturbed
cell  energies $E_k^N$ and cell matrix elelements $\langle
k,N|a_h |k',N'\rangle$, where $a_h$ is a generic annihilation operator
appearing in Eq. (\ref{E:hop}) and $N,N' = L-2,L-1,L,L+1,L+2$. The
form of the perturbative terms is quite similar to those described in
the appendix of Ref. \cite{A:cSCPE} in the case of 1D
superlattices, although more complex, essentially
due to the more complex topology of the structures in
Fig. \ref{F:lattices}. Note indeed that the perturbative 
term in Eq. (\ref{E:hop}) is such that a boson
operator at a given site of a cell may be involved in more than 
one inter-cell hopping, unlike the 1D
case. This (not excessive) increase in complexity is greatly 
rewarded by a very low computational effort, mostly  
required by the complete diagonalization of the cell Hamiltonian 
(\ref{E:cBH}) for a few values of the cell population.
Since the cells of the structures in Fig. \ref{F:lattices} comprise
just two and three sites, we are able to provide satisfactory
quantitative results at a negligible computational cost also for
insulator domains relevant to fillings that would be prohibitive for
brute force numerical methods such as quantum Monte Carlo or density
matrix renormalization group simulations. More in general, the portion
of phase diagram of a superlattice accessible to our perturbative 
technique depends only on the number of sites within a unit cell, and
not by the topology or dimensionality of the cell lattice. We checked the
correctness of our perturbative results against brute force numeric results (Lanczos or quantum Monte Carlo algorithm) on small periodic ladder and kagom\'e lattices (comprising up to 9 cells).
\begin{figure}
\begin{center}
\includegraphics[width=8.5cm]{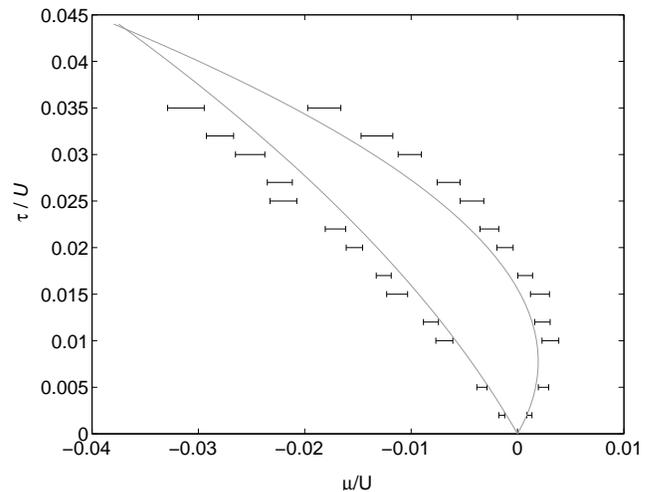}
\caption{\label{F:ladder} Half-filling insulator domain for a TLL
  with intra-cell hopping amplitudes $\tau$ and inter-cell hopping
  amplitudes $\tau/2$. The $2^{\rm nd}$ order cell SCPE result (solid lines) is
  compared to quantum Monte Carlo simulations for a TLL comprising
  50 rungs (errorbars).}
\end{center}
\end{figure}
Our formulas  hold for the most generic superlattice
described by Eqs. (\ref{E:BH})-(\ref{E:hop}). This means that the
intra-cell hopping amplitudes, as well as the energy offsets $v_k$,
may be different from each other. Furthermore, there can be $z/2$
different inter-cell hopping amplitudes, where $z$ is the coordination
number of the cell lattice. Here we assume that $v_k=0$, 
so that the fractional filling insulator domains have a loophole shape
\cite{A:loophole}. Furthermore we assume that all of the intra-cell
hopping amplitudes are equal to $\tau$, and all of the inter-cell hopping
amplitudes are equal to $\tau'$. In these conditions the presence of loophole domains can be related to a simple parameter, i.e. the hopping ratio $\tau'/\tau$, that is basically the perturbative quantity in our expansions. This ratio can be also  connected to the  energy gap in the single particle spectra of the structures under concern, that on 1D structures bears a strict relation to the presence of loopholes, due to the exact mapping between hard-core bosons and free spinless fermions \cite{A:Girardeau,A:loophole}.

In the case of the TLL, the gap in the single particle spectum disappears for $\tau/\tau' \leq 2$, whereas for the TKL it is present for any $\tau \neq \tau'$. Detailed investigations  at half-filling  for the ladder and at $1/3$-filling for the TKL show that the thresholds for the occurrence of the relevant loophole insulator domains differ from those of the single particle gaps. 
Actually, Fig. \ref{F:ladder} shows that, according to both cell SCPE and quantum Monte Carlo simulations, the half-filling loophole domain of the ladder is still present for $\tau/\tau' = 2$, when the single particle gap is not present any more. The comparison between these results also shows that cell SCPE provide quantitatively satisfactory results also when $\tau'$ is a significant fraction of $\tau$, i.e. when the perturbative term is far from being infinitesimal.
Furthermore, both cell SCPE and quantum Monte Carlo simulations for hard-core bosons suggest that the half-filling loophole domain of the ladder persists for $\tau/\tau' < 2$, and disappears when the hopping ratio becomes smaller than a finite quantity smaller than 2. This  partially agrees with the results reported in Ref. \cite{A:Vekua} for a spin ladder that maps onto the hard-core limit of our model  for a suitable parameter choice. 

On the TKL the situation is in some sense reversed, since 
the  loophole domains disappear when a single particle gap is still
present. More precisely, the gap is open for any $\tau'<\tau$, but it must be $\tau'<\tau/c$ for the  loophole insulator domains to occur, where $c = c_1 \approx 2.32$ and $c =c_2\approx 2.87$ according to first and second order perturbative results, respectively. We mention that these values of $c$ are the same for the first three loopholes, relevant to filling $1/3$, $2/3$ and $4/3$. Note that the fact that $c_2 > c_1$ is coherent with our perturbative approach, whose espansion parameter is basically $\tau'/\tau$.
\begin{figure}
\begin{center}
\includegraphics[width=8.5cm]{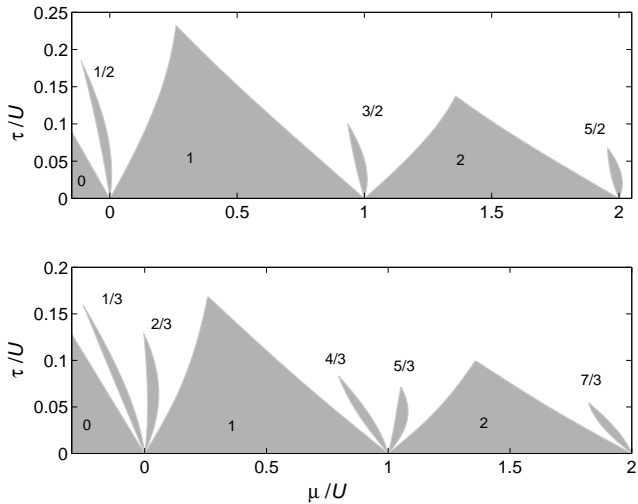}
\caption{\label{F:ph_dgms} Phase diagrams for the structures in Fig.~\ref{F:lattices}.  Top: TLL with $\tau=3 \tau'$. Bottom: TKL with $\tau=6 \tau'$. The fillings of the insulator domains (gray) are also given in the plot.}
\end{center}
\end{figure}

We conclude by commenting Fig.~\ref{F:ph_dgms}, which shows the phase diagrams for the structures under investigation as provided by our second order cell strong coupling expansions. The value of the (fixed) hopping ratio $\tau/ \tau'$ is  3 for the TLL (upper panel) and 6 for the TKL (lower panel). Note that, since the hopping amplitude depends exponentially on the strength of the optical lattice, the above ratios correspond to a relatively small modulation of the potential barriers between neighbouring sites.
We emphasize that these ratios correspond to an arbitrarily small modulation of the potential profile of the superlattice. Indeed, we recall that $\tau \sim e^{-I h}$, $\tau' \sim e^{-I h'}$, where $I$ is the strength of the optical lattice, while $h$ and $h'$ are scaling factors  related to the height of the potential barrier between adjacent sites belonging to the same cell or to different cells, respectively \cite{A:Jaksch}. It is hence clear that a sufficiently large $I$ produces the desired ratio, provided that $h<h'$. This suggests that the fractional filling insulating phase can be always reached if the lattice strength is increased while keeping the potential profile fixed. 
In this respect we emphasize that Figs.~\ref{F:ladder} and \ref{F:ph_dgms} correspond to a slightly different approach, since we assume that $\tau/U$ is varied while keeping  the  ratio $\tau/ \tau'$  fixed. This can be done by adjusting the setup and intensity of the laser beams giving rise to the optical superlattice, at least in principle \cite{A:Roth03,A:Peil,A:loophole}. However it may prove more convenient to keep the beam configuration fixed (i.e. to fix $\tau$ and $\tau'$), while varying the interaction strength $U$, e.g. via Feshbach resonances \cite{A:Theis}.


\begin{thebibliography}{10}
\bibitem{N:ovvio}
Since the cells are identical, the energy offsets $v_k$ and the hopping
  amplitudes $T_{h k}$ do not depend on the cell index.
\end{thebibliography}
\end{document}